\documentclass{aa}
\usepackage{amssymb}
\usepackage{graphicx}
\usepackage{txfonts}
\begin{document}
   \title{On the influence of ram-pressure stripping on the star formation of simulated spiral galaxies}

   \author{T. Kronberger \inst{1}
          \and
           W. Kapferer \inst{1}
          \and
           C. Ferrari \inst{1}
          \and
           S. Unterguggenberger \inst{1}
          \and
           S. Schindler \inst{1}
           }

   \offprints{T. Kronberger, \email{Thomas.Kronberger@uibk.ac.at}}

   \institute{Institute for Astro- and Particle Physics, University
   of Innsbruck, Technikerstr. 25, A-6020 Innsbruck, Austria
   }

   \date{-/-}

% \abstract{}{}{}{}{}
% 5 {} token are mandatory

  \abstract
   {}
   {We investigate the influence of ram-pressure stripping on the star formation and the mass
   distribution in simulated spiral galaxies. Special emphasis is put on the question where
   the newly formed stars are located. The stripping radius from the simulation is compared
   to analytical estimates.}
   {Disc galaxies are modelled in combined N-body/hydrodynamic simulations (GADGET-2) with prescriptions for
   cooling, star formation, stellar feedback, and galactic winds. These model galaxies move through
   a constant density and temperature gas, which has parameters comparable to the intra-cluster medium (ICM) in the outskirts of a galaxy cluster
   (T=3 keV $\approx$ 3.6$\times$10$^7$ K and $\rho$=10$^{-28}$ g/cm$^3$). With this numerical setup we analyse the influence
   of ram-pressure stripping on the star formation rate of the model galaxy.}
   {We find that the star formation rate is significantly enhanced by the ram-pressure effect (up to a factor of 3). Stars form in the
   compressed central region of the galaxy as well as in the stripped gas behind the galaxy. Newly formed
   stars can be found up to hundred kpc behind the disc, forming structures with sizes of roughly 1 kpc in diameter and with
   masses of up to 10$^7$ M$_{\odot}$. As they do not possess a dark matter halo due to their formation history, we name them 'stripped baryonic dwarf'
   galaxies. We also find that the analytical estimate for the stripping radius from a Gunn \& Gott (1972) criterion is in good agreement with
   the numerical value from the simulation. Like in former investigations, edge-on systems lose less gas than face-on systems and the resulting
   spatial distribution of the gas and the newly formed stars is different.}
   {}


   \maketitle
%
%________________________________________________________________

\section{Introduction}
Many issues of galaxy formation and evolution can be understood
within the framework of the successful Cold Dark Matter (CDM)
theory. But besides the hierarchical assembly of galaxies via
mergers, other interaction phenomena are important in the cluster
environment. Different types of such interactions are discussed:
galactic winds, ram-pressure stripping, galaxy-galaxy interactions
and outflows from Active Galactic Nuclei (AGN). Most of these
processes can occur, with some principal differences, also in the
field environment. Ram-pressure stripping, however, is uniquely
linked to the presence of a surrounding medium that acts by its ram
pressure. The space between the galaxies in a galaxy cluster is
filled with a hot ($\sim 10^8$ K), thin ($\sim 10^3$ ions/m$^3$)
plasma, the so called intra-cluster medium (ICM). Theoretically, the
interaction between galaxies in the cluster and the ICM was already
described by Gunn \& Gott (1972). They argued that galaxies lose
material if the force due to ram-pressure stripping exceeds the
restoring gravitational force of the galaxy. As a consequence, a
stripping radius in ram-pressure affected galaxies can be defined,
outside which the inter-stellar medium (ISM) cannot be prevented
from being stripped by the galactic gravitational potential. Another
analytical estimate was proposed by Mori \& Burkert (2000) where the
thermal ISM pressure is compared with the external ram pressure.
Later it was also found observationally that ram pressure acts on
several galaxies in the Virgo cluster (e.g. Cayatte et al. 1990;
Kenney et al. 2004; Vollmer et al. 2004) and in the Coma cluster
(Bravo-Alfaro et al. 2000, 2001). Those galaxies, which are subject
to ram-pressure stripping, lose parts of their inter-stellar medium
to the ICM. As this gas has been enriched with heavy elements by the
stars within the galaxy, the chemical abundance of the surrounding
intra-cluster medium will be affected (Schindler et al. 2005;
Domainko et al. 2006; Kapferer et al. 2007). These heavy elements
can then be found in X-ray observations.

Besides this treatment in cluster-scale simulations, several groups
have already studied numerically the effect of ram-pressure
stripping on individual galaxies. The stripping radius with respect
to acting ram pressure and galaxy properties was investigated (e.g.
Abadi et al. 1999, Vollmer et al. 2001, Roediger \& Hensler 2005)
and a reasonable agreement between the analytically estimated
stripping radius and the stripping radius from numerical simulations
was found. Further galaxy-scale numerical simulations for
ram-pressure stripping were presented by Quilis et al. (2000), Mori
\& Burkert (2000), Toniazzo \& Schindler (2001), Schulz \& Struck
(2001), and Roediger \& Br\"uggen (2006,2007). Very recently
J{\'a}chym et al. (2007) used an N-body/SPH code to investigate the
influence of a time-varying ram pressure on spiral galaxies.

In contrast to these papers we focus on the effect of ram-pressure
stripping on the star formation rate (SFR) of the galaxy, rather
than on the mass loss. Nevertheless, for completeness we also
analyse the mass loss and the resulting disc structure. The combined
effects of ram-pressure stripping and tidal interactions on
simulated cluster spirals is investigated in detail in Kapferer et
al. (2007b).

This paper is organized as follows: In Sect. 2 we present the
numerical method and the simulation setup used for this work. In
Sect. 3 the results are presented which are compared to observations
in Sect. 4. We end with a summary of the main conclusions in Sect.
5.

\section{The numerical setup}
The simulations were carried out with the N-body/SPH code GADGET-2
developed by V. Springel (see Springel 2005 for details). This code
treats the gas of the galaxies and the ICM hydrodynamically via
smoothed particle hydrodynamics (SPH, Gingold \& Monaghan 1977; Lucy
1977) while the collisionless dynamics of the dark matter and the
stellar component is calculated using an N-body technique.
Prescriptions for cooling, star formation (SF), stellar feedback,
and galactic winds are included as described in Springel \&
Hernquist (2003). As we focus here on the effects of ram-pressure
stripping on the star formation rate of the model galaxy, we
summarise the main properties of the star formation model presented
in  Springel \& Hernquist (2003). In this formalism it is not
attempted to spatially resolve star formation in the ISM but to
describe star formation in a statistical manner. An SPH particle
represents a certain region in the ISM with given values for the
spatially averaged hydrodynamic quantities. Stars are assumed to
form from cold gas clouds on a characteristic timescale t$_{\star}$
and a certain mass fraction $\beta$ is instantaneously released due
to supernovae from massive stars ($>$ 8 M$_{\odot}$). The simple
model for the time evolution of the stellar density $\rho_{\star}$
reads:

\begin{equation}
\frac{d\rho_{\star}}{dt}=\frac{\rho_{c}}{t_{\star}}-\beta\frac{\rho_{c}}{t_{\star}}=
(1-\beta)\frac{\rho_{c}}{t_{\star}}
\end{equation}

\noindent where $\rho_{c}$ is the density of the cold gas. As in
Springel \& Hernquist (2003) we adopt $\beta$ = 0.1 assuming a
Salpeter IMF with a slope of -1.35 in the limits of 0.1 M$_{\odot}$
and 40 M$_{\odot}$. In addition to the mass injection also energy is
released to the ISM by the supernovae. This feedback energy heats
the ambient gas and cold clouds can be evaporated inside hot bubbles
of supernovae. All these assumptions lead to a self regulating star
formation, where the growth of cold clouds is balanced by supernova
feedback of the formed stars. The mass budget for the hot and cold
gas densities ($\rho_{h}$ and $\rho_{c}$, respectively) due to star
formation, mass feedback, cloud evaporation and growth of clouds due
to radiative cooling can then be written as

\begin{equation}
\frac{d\rho_{c}}{dt}=-\frac{\rho_{c}}{t_{\star}}-A\beta\frac{\rho_{c}}{t_{\star}}+
\frac{1-f}{u_{h}-u_c}\Lambda_{net}(\rho_h,u_h),
\end{equation}
\noindent and
\begin{equation}
\frac{d\rho_{h}}{dt}=\beta\frac{\rho_{c}}{t_{\star}}+A\beta\frac{\rho_{c}}{t_{\star}}-
\frac{1-f}{u_h-u_c}\Lambda_{net}(\rho_h,u_h)
\end{equation}

\noindent where \textit{f} represents a factor to differentiate
between ordinary cooling (f=1) and thermal instability (f=0),
$\Lambda_{net}$ is the cooling function (Katz et al., 1996), and $A$
is the efficiency of the evaporation process. The onset of the
thermal instability and hence of the star formation is set by a
simple density threshold criterion, motivated by observations
(Kennicutt 1989). Thus f$=$0 for densities larger than a certain
threshold density $\rho_{th}$ and otherwise f$=$1.

The mass loss of the galaxy due to galactic winds, $\dot{M}_w$, is
assumed to be proportional to the star formation rate $\Psi_{SFR}$,
i.e. $\dot{M}_w=\eta\Psi_{SFR}$ with $\eta=2$, consistent with the
observations of Martin (1999). Additionally, the wind contains a
fixed fraction $\chi$ of the supernova energy, which is assumed to
be $\chi$=0.25. For further details and a description of the
numerical implementation of this model we refer to Springel \&
Hernquist (2003). This wind model is a purely phenomenological one,
that lacks a physical foundation. Nevertheless, the model is
constrained by observational data, as far as possible, and should
therefore be able to give plausible results.

The initial conditions were built according to Springel et al.
(2005), based on the analytical work of Mo et al. (1998). The model
galaxy, which we use, represents a Milky-Way type spiral galaxy. The
total mass of the model galaxy is 1.09$\times$10$^{12}$ $M_{\sun}$,
where the initial total gas mass is 6.8$\times$10$^{9}$ $M_{\sun}$.
Compared to previous work (Kapferer et al. 2005, 2006; Kronberger et
al. 2006, 2007), where we used a similar sized galaxy, we have
significantly increased the number of particles, increasing the mass
resolution of the simulation. Important numerical quantities, such
as the particle numbers, are summarized in Table
\ref{galaxyproperties_resolution}. For the intra-cluster medium we
use one million SPH particles, which we distribute homogeneously
over a volume of 1 Mpc$^3$ with a mass density of $1\times10^{-28}$
g/cm$^3$ and a constant temperature of 3 keV ($\sim 3.6\times10^{7}$
K). Such a simplified ICM distribution allows to study the effects
of ram-pressure stripping in a clean way, i.e. with as little
degeneracies of different effects as possible. Additional effects
from varying density and temperature in the ambient medium will be
investigated in an upcoming work.

\begin{table}
\begin{center}
\caption[]{Particle numbers and mass resolution used for the model
galaxy. Additionally the circular velocity of the halo at
$\rm{r}_{200}$ and the disc scale length of the initial conditions
are given.}
\begin{tabular}{c c c}
\hline \hline & particle number & mass resolution \cr & &
[$M_{\sun}$/particle] \cr\hline Dark matter halo & 300000 &
$3.5\times10^{6}$ \cr Disk collisionless & 200000 &
$1.0\times10^{5}$ \cr Gas in disk & 200000 & $3.4\times10^{4}$ \cr
ICM & 1000000 & $1.4\times10^{6}$\cr\hline
\end{tabular}
\label{galaxyproperties_resolution}
\end{center}
\end{table}

Starting from these initial conditions we calculate the evolution of
the model galaxy moving with a constant velocity of 1000 km/s
through the ambient medium for 1 Gyr. A galaxy typically will not
move through a homogeneous ICM for 1 Gyr. Therefore, to keep our
results comparable to observations of galaxies in the outskirts of
galaxy clusters, we study primarily the first few hundred Myr. We
calculated two different interaction geometries, one in which the
galaxy flies face-on through the ambient medium and one in which the
ram-pressure acts edge-on. In order to avoid numerical artifacts we
use for the investigation of ram-pressure stripping a model galaxy
after 2 Gyr of isolated evolution. After that period of time
self-regulated star formation is present and the gas distribution
has settled from the simple analytic initial condition to an
equilibrium distribution with a spiral structure.

To study possible resolution effects, originating from the different
mass resolution of ICM and ISM particles, we carry out a simulation
with 10$^7$ ICM particles. We found no significant differences to
the lower resolution runs concerning the results presented below for
the star formation rate or the mass distribution.

The SPH method has some intrinsic shortcomings that might affect the
results to some extend. One is the resolution of instabilities and
turbulence (e.g. Agertz et al. 2007). The shear flows present in
systems affected by ram-pressure stripping can lead to the
development of Kelvin-Helmholtz instabilities (e.g. Roediger et al.
2006), which are not resolved in our setup. Instabilities should
induce turbulence and fragmentation and therefore enhance star
formation which is consistent with our result. A similar study with
a Eulerian code is a desirable future work to investigate the
importance of hydrodynamic instabilities in this issue.

\section{Results}
In the following we will first analyse the mass loss of the model
galaxy due to ram-pressure stripping. Then we will present the
results concerning the star formation rates and the distribution of
the newly formed stars, within the adopted hydrodynamic and
star-formation model.

\subsection{The stripping radius and the mass distribution inside the stripped galaxy}
The Gunn \& Gott (1972) criterion, according to which galaxies lose
material if the force due to ram-pressure stripping exceeds the
restoring gravitational force of the galaxy, implies the existence
of a stripping radius. This is the distance from the galactic
centre, outside which the galactic gravitational potential cannot
prevent the ISM from being stripped. The position of this stripping
radius depends on the mass distribution in the galaxy and the
external pressure exerted by the ICM. Assuming a double exponential
density distribution of the stars and the gas in the disc, the
expression for the stripping radius, R$_{strip}$, reads (Domainko et
al., 2006):

\begin{equation}
\frac{R_{strip}}{R_0}=0.5\times
ln\left(\frac{GM_{star}M_{gas}}{v_{gal}^2\rho_{ICM}2\pi
R_0^4}\right). \label{gandg}
\end{equation}

\noindent Here M$_{star}$ and M$_{gas}$ are the mass of the stellar
and gaseous disc, respectively, v$_{gal}$ denotes the velocity of
the galaxy relative to the ICM, $\rho_{ICM}$ is the density of the
surrounding ICM, and R$_0$ is the radial scale length of the disc.
The validity of this approximation was already shown using numerical
simulations and by comparison with Virgo cluster galaxies by Abadi
et al. (1999).

\begin{figure}
\begin{center}
{\includegraphics[width=\columnwidth]{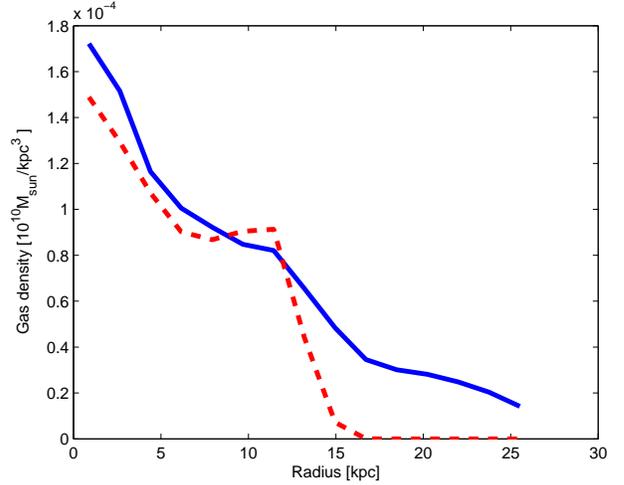}} \caption{Radial
gas density profile for the isolated model galaxy (solid line) and
the stripped model galaxy (dashed line), which moves face-on through
the ICM. The distribution is shown after 100 Myr of ram pressure
acting. A drop of the ISM density at $\sim$ 13 kpc is clearly
visible.} \label{stripping_radius1}
\end{center}
\end{figure}

We also find a fair agreement between the analytical value and the
stripping radius in our simulation. From Eq. \ref{gandg} we
calculate a theoretical value of 13.2 kpc. In Fig.
\ref{stripping_radius1} the gas density profile of the stripped
model galaxy (dashed line), which moves face-on through the ICM, is
compared to the isolated one (solid line) at the same evolutionary
time. The ISM distribution is shown after 100 Myr of ram pressure
acting. There is a clear drop in the density at a radius of
r$\sim$13 kpc visible, which corresponds well to the analytically
derived stripping radius. At radii larger than 15 kpc the gas is
almost completely stripped. While in the simple theoretical model
the gas distribution inside the stripping radius is not affected, it
adjusts in the hydrodynamic simulation to the modified mass
distribution and pressure gradients and changes due to the mass loss
caused by galactic winds. At radii r$<$10 kpc the gas density lies
below the density of the isolated galaxy, while near to the
stripping radius the density increases with respect to the isolated
one and forms a peak. In the central regions of the ram-pressure
affected disc the star-formation rate is enhanced (see Sect.
\ref{sfr_section}), which depletes the gas reservoir there due to
the conversion of gas to stars and a subsequently enhanced galactic
wind. Once established this density distribution remains
qualitatively in this shape, while quantitatively it changes with
time due to the acting galactic wind.

\begin{figure}
\begin{center}
{\includegraphics[width=\columnwidth]{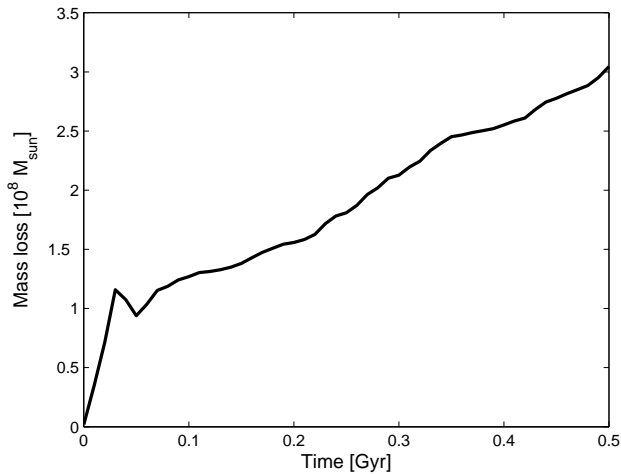}}
\caption{Integrated total mass loss due to ram-pressure stripping
and galactic winds of the stripped model galaxy, which moves face-on
through the ICM. Note that only the gravitationally unbound gas is
plotted. Even a decrease of the mass loss is possible, if gas
enters, for example, the slipstream of the galaxy, as visible after
$\approx$50 Myr} \label{massloss}
\end{center}
\end{figure}

The total gas mass loss due to ram-pressure stripping and galactic
winds is shown in Fig. \ref{massloss} for the first 500 Myr. For
this figure we take only the gravitationally unbound gas into
account, i.e. the gas which has a positive total energy. Note that
this mass loss is a superposition of two processes, ram-pressure
stripping and galactic winds. There is an interplay between both
effects, as ram-pressure stripping changes the star formation rate,
which in turn changes the galactic wind mass loss. As many authors
have already investigated the mass loss due to ram-pressure
stripping and compared the numerical results to analytical estimates
(e.g. J{\'a}chym et al. 2007) we do not study the isolated effect of
ram-pressure stripping here. However, we point out once more, that
the stripping radius in our simulations is comparable to analytical
estimates, as found by various authors. In the time after the
stripping radius becomes clearly visible, galactic winds should be
the dominating outflow mechanism. The mass loss stays then roughly
constant over time. Note, however, that the ambient medium acts on
the galactic wind by its pressure and that the mass of the gas
actually lost by the galactic wind is hence smaller than expected
from the star formation rate. Such a suppression of galactic winds
has already been discussed in a different context by Schindler et
al. (2005).

Additionally we investigate the spatial distribution of the stripped
gas. We find that after 100 Myr 13.7 \% (9.3$\times$10$^8$
M$_{\odot}$) of the initial ISM mass is in the wake of the galaxy.
We define this to be the region more than 6 times the scale height
away from the disc, which is for our model galaxy 1.3 kpc. This
distance should be significantly larger than the half-light radius
of the galaxy, and therefore observationally be considered as not
belonging to the galaxy. After 200 Myr, this amount of stripped gas
increases to 31.7 \% (2.2$\times$10$^9$ M$_{\odot}$) of the initial
ISM mass. However, material caught in the slipstream of the moving
galaxy does not feel the ram pressure of the ICM and a significant
proportion will fall back onto the disk. After 500 Myr about 10 \%
(6.8$\times$10$^8$ M$_{\odot}$) of the initial ISM mass are in the
wake of the galaxy and gravitationally unbound.

We perform the same analysis for the case in which the model galaxy
moves edge-on through the medium. In this case no clear stripping
radius can be identified. The gaseous disc is compressed and
distorted but the mass loss is lower than for the face-on galaxy.
After 100 Myr there is only 0.9 \% (6.1$\times$10$^7$ M$_{\odot}$)
of the initial ISM mass in the wake of the galaxy. This fraction
increases to 8.6 \% (5.8$\times$10$^8$ M$_{\odot}$) after 200 Myr
and finally after 500 Myr 6.3 \% (4.3$\times$10$^8$ M$_{\odot}$) of
the initial ISM mass is in the wake of the galaxy and
gravitationally unbound. Also in the edge-on case we find
re-accretion in the slipstream of the galaxy.

\subsection{The impact on star formation}\label{sfr_section}
\begin{figure}
\begin{center}
{\includegraphics[width=\columnwidth]{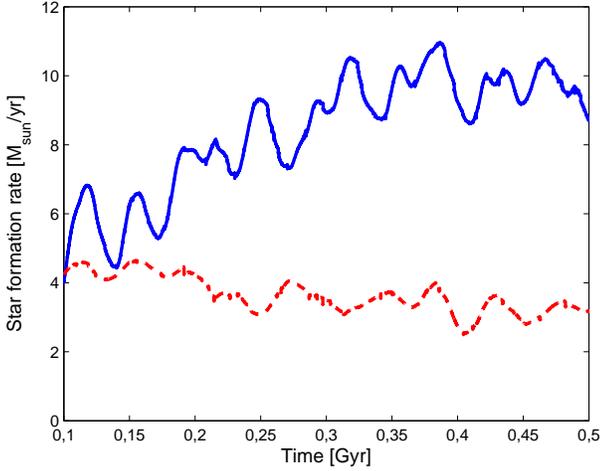}} \caption{Total
star formation rate as a function of time for an isolated model
galaxy (red, dashed line) and for a ram-pressure stripped model
galaxy (blue, solid line), which moves face-on through the ICM. Note
that the increase of the star-formation rate is the physically
relevant information. The apparent oscillations are due to the
numerical implementation of the star-formation model.} \label{sfr}
\end{center}
\end{figure}

\begin{figure}
\begin{center}
{\includegraphics[width=\columnwidth]{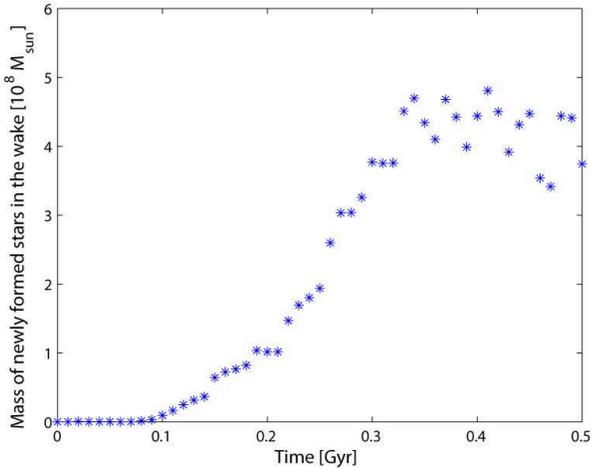}} \caption{Mass
of stars formed in the wake of the model galaxy as a function of
time. The ram pressure acts face-on. The scatter in the plot is a
consequence of stars falling back into the gravitational potential
towards the centre of the disc.} \label{stars_wake}
\end{center}
\end{figure}

There has been a lot of discussion whether ram-pressure stripping
acting on a galaxy enhances or decreases the star formation rate
(e.g. Evrard 1991; Fujita et al. 1999). With our simplified setup,
that studies the influence of a homogeneous medium, we find a clear
enhancement, within the adopted star formation model. In Fig.
\ref{sfr} we plot the total star formation rate as a function of
time for an isolated model galaxy and for a ram-pressure stripped
model galaxy, which moves face-on through the ICM. Over 500 Myr the
star formation rate is always significantly (up to a factor of 3)
higher than in the isolated galaxy. Most of the new stars form in
the disc, mainly in the compressed central region. However, star
formation is also present in the stripped material in the wake of
the galaxy. These stars are not gravitationally unbound but can
spread over a large volume in the galactic halo. We find newly
formed stars up to hundred kpc behind the plane of the disc. As
these stars do not feel the ram pressure of the surrounding ICM,
they fall back into the gravitational potential towards the centre
of the disc. The kinetic energy they gain hereby is sufficient that
they partly cross the galactic plane and enter the space in front of
the disc. In the ram-pressure affected galaxy about twice as many
stars are formed as in the isolated galaxy in the time while
ram-pressure stripping is acting. After 200 Myr of acting ram
pressure the total mass of newly formed stars in this period is
1.02$\times$10$^8$ M$_{\odot}$ in the wake of the galaxy, while in
the same time in the disc 9.5$\times$10$^8$ M$_{\odot}$ form. (The
wake of the galaxy is defined, as before, to be more than 6 times
the scale height away from the disc. The disc region itself extends,
by our definition, to four times the stellar disc scale length in
radial direction and twice the scale height in the direction
perpendicular to the disc.) Note that these masses also depend on
the chosen IMF, as this regulates the amount of massive stars that
end instantaneously (i.e. in less than a typical numerical time
step) as supernovae. Therefore the mass of the newly formed stars
does not include the high-mass stars ($>$ 8 M${_\odot}$). In Fig.
\ref{stars_wake} we show the mass of the newly formed stars as a
function of time. The scatter in the plot is a consequence of stars
falling back into the gravitational potential towards the centre of
the disc.

\begin{figure}
\begin{center}
{\includegraphics[width=\columnwidth]{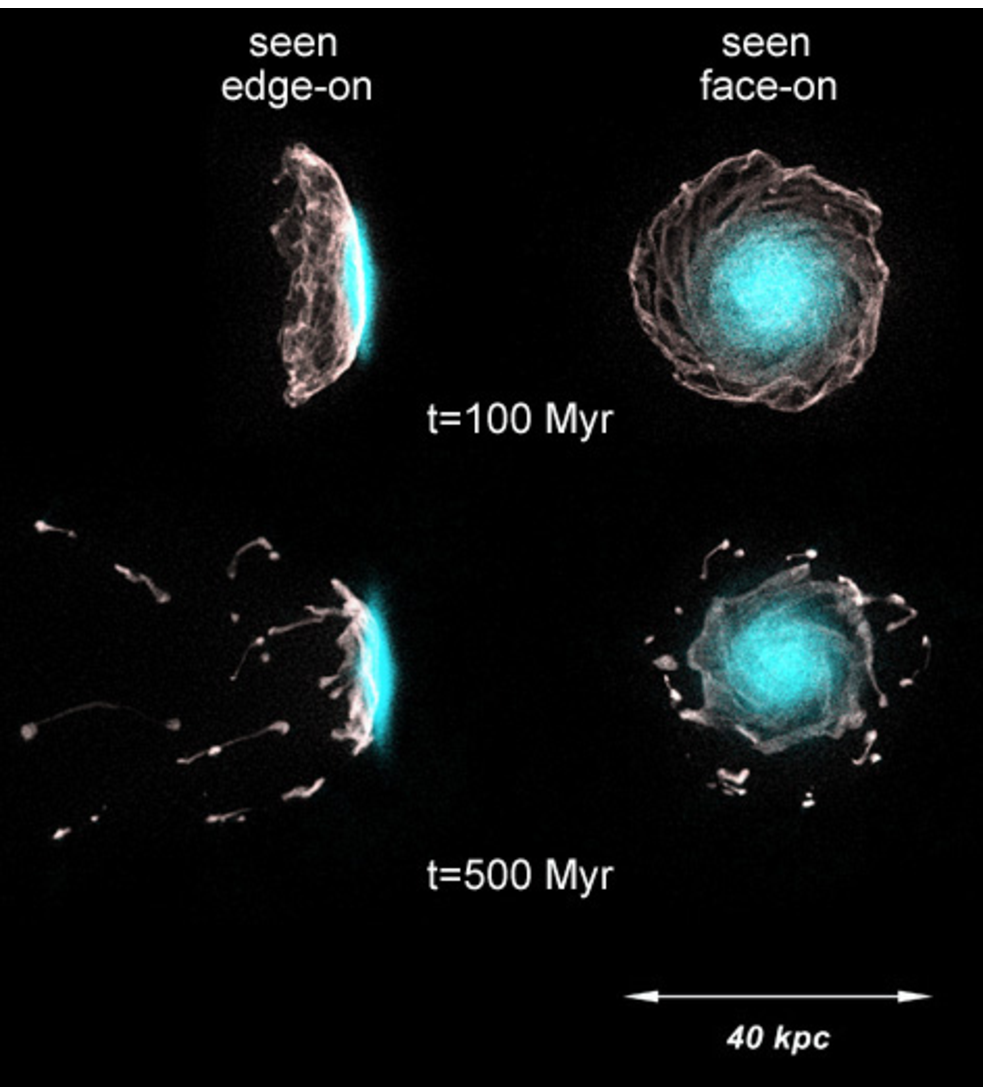}}
\caption{Distribution of the gas (white) and the newly formed stars
(turquoise) for a ram-pressure affected galaxy, which moves face-on
through the ICM, after 100 Myr (top) and after 500 Myr (bottom),
seen face-on and edge-on.} \label{all}
\end{center}
\end{figure}

\begin{figure}
\begin{center}
{\includegraphics[width=\columnwidth]{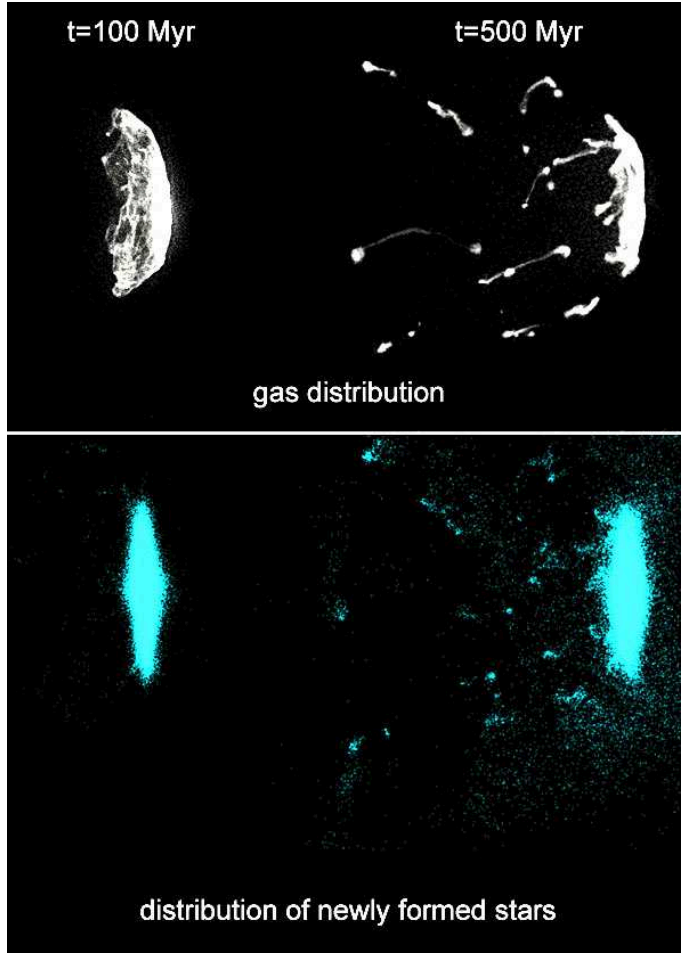}}
\caption{Distribution of the gas (white, top) and the newly formed
stars (turquoise, bottom) seen edge-on for a ram-pressure affected
galaxy, which moves face-on through the ICM, after 100 Myr (first
column) and after 500 Myr (second column).} \label{comp}
\end{center}
\end{figure}

The distribution of the gas and the newly formed stars is presented
in Figs. \ref{all} and \ref{comp} for the ram-pressure affected
galaxy, which moves face-on through the ICM. Note that newly formed
stars are here defined to be stellar particles formed after the
beginning of the simulation. After 100 Myr the stripping is already
active and a stripping radius is visible. A wake, however, has not
yet formed, and the star formation mainly takes place in the central
part of the galaxy, where the gas is compressed by the ram pressure
of the surrounding ICM. After 500 Myr a wake has formed with dwarf
galaxy sized sub-structures (diameter $\sim$1 kpc, masses up to
10$^7$ M$_{\odot}$ in gas and newly formed stars). These objects
formed by initial compression and subsequent cooling of already
slightly overdense regions in the disc (parts of spiral arms). This
formation scenario is different from the gravitational collapse, as
discussed for tidal dwarf galaxies in tidally interacting systems.
Note that due to their formation history from stripped material,
these structures do not have a dark matter halo. We therefore term
these structures 'stripped baryonic dwarf' galaxies. Most of these
structures eventually fall back onto the galactic disc on timescales
of $\leq$ 1 Gyr. In more realistic systems with a gravitational
cluster potential and interactions with other galaxies the fate of
these substructures might be different. Especially the ram pressure
of the ICM usually changes direction, as not all orbits are purely
radial. Kapferer et al. (2007b) show that such structures also form
when galaxy - galaxy interactions take place in an ambient medium.
They are then distributed over much larger distances compared to the
case presented here. It has to be further investigated, whether
these objects can then form virialised, long-term stable structures.
In Fig. \ref{comp} we show gas and newly formed stars separately, in
order to show their distribution more clearly. As stars are
collisionless their distribution differs from that of the gas, the
small substructures are, however, visible in both distributions.
From the stripped gas only around 25\% is heated to temperatures
above 1$\times$10$^7$ K after 500 Myr of acting ram pressure, and
hence visible in X-rays. Especially in the dense knots, where stars
are formed, the temperature remains significantly lower.

\begin{figure}
\begin{center}
{\includegraphics[width=\columnwidth]{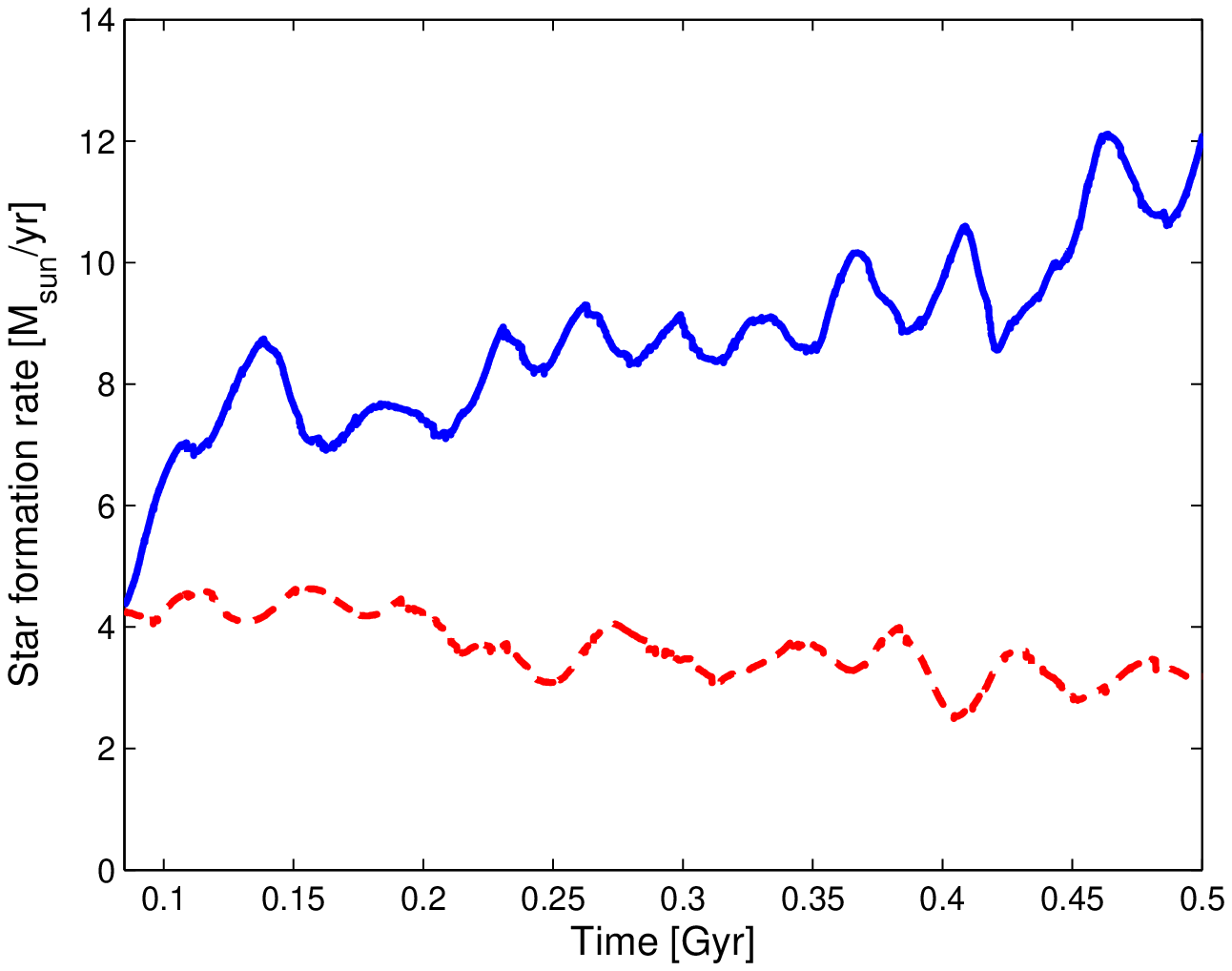}} \caption{Total
star formation rate as a function of time for an isolated model
galaxy (red, dashed line) and for a ram-pressure stripped model
galaxy (blue, solid line), which moves edge-on through the ICM. Note
that the increase of the star-formation rate is the physically
relevant information. The apparent oscillations are due to the
numerical implementation of the star-formation model.}
\label{sfr_eo}
\end{center}
\end{figure}

The situation is very similar in the edge-on case, although the
resulting mass distribution is different. In Fig. \ref{sfr_eo} the
total star formation rate as a function of time is shown for an
isolated model galaxy and for a ram-pressure stripped model galaxy,
which moves edge-on through the ICM. The enhancement of the star
formation rate is here even stronger than in the face-on case. These
new stars are, however, mainly formed in the central region of the
disc. The gas from the outskirts of the galactic disc is pushed to
the central region, enhancing there significantly the density of the
cold gas and hence the star formation rate.

From the results presented in this section we would therefore expect
especially in galaxy cluster mergers an increased star formation
activity in the merger region, because the galaxies encounter there
a rapid rise of the external ram pressure. As a lot of gas from the
galaxy is lost by ram-pressure stripping or converted to stars, the
cold gas reservoir is constantly depleted. Hence, the star formation
rate first increases and eventually decreases.

\section{Comparison of the results to observations}
An evolution of the cluster population with redshift has been
pointed out since the late 70's. The fraction of star-forming and
post-star-forming systems has been proven to significantly increase
with $z$, going from $\geq$30\% at $z{\sim}0.3-0.5$ (Dressler et al.
1999) to $\sim$1-2\% in the local Universe (e.g. Dressler 1987). In
agreement with this tendency, in 1978 Butcher \& Oemler reported a
strong evolution from bluer to redder colours in cluster galaxies,
detecting an excess of blue objects at $z$=0.5 with respect to lower
redshift systems ( ``Butcher \& Oemler effect'').  In order to link
the observed evolution of cluster members to the hierarchical growth
of structures, we still need to understand the specific importance
of each of the proposed physical mechanisms acting on the star
formation activity and on the morphology of galaxies (i.e. galaxy
mergers and interactions: e.g. Lavery \& Henry 1988 -- ram-pressure
stripping: Gunn \& Gott 1972 -- ``harassment'': e.g. Moore et al.
1998 -- ``strangulation'' or ``starvations'': e.g. Larson et al.
1980).

Quantifying the evolution of SF in cluster members, however, is
extremely difficult, since the fraction of star-forming objects
varies from one cluster to another also in the same redshift range
(see Poggianti 2004 and references therein). It is thus essential to
study simultaneously the evolution of galaxies as a function of
redshift and of the properties of their host cluster. Ferrari et al.
(2005) have for instance proven that the fraction of star-forming
and post-star-forming galaxies in the low redshift ($z\simeq$0.09)
cluster A3921 is comparable to those measured at intermediate
redshifts. A3921 is a merging system (Ferrari et al. 2005; Belsole
et al. 2005), and a detailed comparison of optical and radio
observations revealed that most of the star-forming galaxies are
located in the collision region of the two interacting
sub-structures (Ferrari et al. 2005, 2006). This suggests that the
high fraction of star-forming objects in A3921 is very likely
related to its dynamical state.

Other analyses agree with this result, suggesting that SF is either
triggered by cluster mergers (e.g. Coma: Caldwell et al. 1993 --
A3562: Bardelli et al. 1998; Miller 2005 -- A2125: Dwarakanath \&
Owen 1999; Owen et al. 2005 -- A2255: Miller \& Owen 2003), or
quenched after a starburst phase, leaving young and strong
poststarburst galaxies (e.g. Coma: Poggianti et al. 2004). Other
studies (e.g. A168: Tomita et al. 1996 -- A3528: Bardelli et al.
2001) did not show any correlation between cluster mergers and
on-going SF. All these results suggest that cluster collisions are
able to affect the internal properties of galaxies, and in
particular their star formation history, possibly inducing an
initial starburst phase, that consumes most of the gas in the
galaxy, creating a subsequent decrease in the star formation rate of
cluster members. Note that, among the previously mentioned clusters,
those systems that are observed $\sim$0.2 Gyr before/after the
closest cores encounter show increased SF (e.g. A3921: Kapferer et
al. 2006b -- A2125: Miller et a. 2005 -- A2255: Sakelliou \& Ponman
2006), contrarily to A168, that instead is as post-merger cluster
observed $\sim$0.6 Gyr after the merging event (Tomita et al. 1996).
The SFR of galaxies would therefore be increased when they are
located in the collision region of a cluster in the very central
phases of merging ($\lesssim$ 0.5 Myr after the closest cores
encounter). One has however to keep in mind that the observational
results presented above are based on observations at different
wavelengths (mostly optical and radio), that therefore trace SF
episodes on different timescales. Homogeneous multi-wavelength
studies of larger samples of clusters at different phases of merging
are thus required to confirm the simple scenario about the effect of
ram pressure on SF presented above.

In agreement with the results of our simulations, it has been
suggested that the enhanced SF activity of the galaxies located in
the collision region of merging clusters is due to dynamical ISM/ICM
interactions (e.g. Poggianti et al. 2004; Ferrari et al. 2006).
Actually, the ISM compression due to ICM ram pressure is most
efficient where the density of the ICM and the relative velocity
between the galaxy and the ICM reach their maximum, i.e. around the
ICM compression bar originated by cluster mergers. Our numerical
results show that, as expected, ram pressure is able to enhance the
SFR of galaxies for several hundred Myr. In our idealised setup we
do not find a significant drop in the SFR due to the depletion of
the cold gas reservoir within our simulation time. However, in more
realistic systems, where the galaxy encounters a density gradient
and moves towards a denser cluster centre, we expect that the gas is
consumed more rapidly and that the SFR drops on shorter timescales.

The dependence of SF on the dynamical state of clusters could be one
of the key ingredients to explain the observed difference in the
fraction of star-forming galaxies in similar redshift systems.
Additionally, this dependence could shed more light on the origin
both of the Butcher \& Oemler effect, and of the spectral evolution
of the galaxies from higher to lower redshift: we know that in the
hierarchical scenario of structure formation the merging rate
between sub-clusters increases with redshift. If the net role of
mergers is to trigger SF, this could be one of the possible
explanations of the excess of blue galaxies detected a $z\sim$0.5
with respect to lower redshifts. Also the lower fraction of active
cluster galaxies in the present Universe could be related to the
lower fraction of merging clusters, and to the fact that member
galaxies of local clusters have already experienced previous cluster
interactions, being therefore gas-poor compared to their high-$z$
counterparts.

Ram pressure {\it alone} cannot explain the observed variation with
density of the SFR in local cluster members. It has been proven
(e.g. Balogh et al. 2004; Haines et al. 2007) that the equivalent
width of the H$\alpha$ emission line in star-forming galaxies does
not depend strongly on the local density, while the fraction of
star-forming objects does. This means that, in order to preserve the
shape of the EW(H$\alpha$) distribution, the physical mechanism(s)
responsible for switching-off SF when a galaxy enter a cluster must
act on short timescales ($<$1 Gyr). Our simulations indicate that,
in about 1 Gyr, ram-pressure stripping and galactic winds remove
only $\sim$13\% of the gas mass. This indicates that in nature
additional processes are at work, which are not modelled in our
simulations.

Based on our results (see top panel of Fig. \ref{comp}), the
existence of very extended ($\sim$ 100 kpc) tails of gas, possibly
associated to intracluster star-forming regions (e.g. Gavazzi et al.
2001; Wang et al. 2004; Yoshida et al. 2004; Oosterloo \& van Gorkom
2005; Sun \& Vikhlinin 2005; Sun et al. 2006, 2007), cannot {\it
only} be due to ram-pressure effects. They are possibly related to
ram-pressure stripping of tidal tails and gaseous bridges created
during galaxy-galaxy interactions (Kapferer et al. 2007b). The work
presented in this paper is on the other hand able to explain the
existence of extraplanar star-forming regions in several cluster
late-type galaxies, that show evidence for ram-pressure stripping,
but no extended tails of intracluster star-forming regions or clear
signatures of interactions with other galaxies (e.g. Cortese et al.
2004; Kenney et al. 2004; Crowl et al. 2005; Cortese et al. 2007).
Ram-pressure stripping acting on both isolated and interacting
galaxies can finally be one of the possible physical mechanisms
responsible for the formation of the observed intracluster stars
(e.g. Freeman et al. 2000; Arnaboldi et al. 2003).

\section{Summary and conclusions}
We have investigated the influence of ram-pressure stripping on the
star formation and the resulting mass distribution in simulated
spiral galaxies. We used a simplified model, where the ICM is
distributed homogeneously and with a constant temperature, in order
to isolate the effects caused by ram pressure.

\begin{itemize}
\item We found that ram-pressure stripping enhances the
star-formation rate by up to a factor of 3 over several hundred Myr.
In total the mass of newly formed stars is about two times higher
than in an isolated galaxy after 500 Myr of ram pressure acting.

\item The new stars are mainly formed in the central parts of the
disc but a significant fraction forms also in the wake of the
galaxy. As they are collisionless they do not feel the ram pressure
and remain bound to the galaxy.

\item Sub-structures form in the wake of the galaxy from compressed
spiral arms of the original galaxy. These irregular shaped
structures have sizes of roughly 1 kpc in diameter and harbour newly
formed stars. We term these structures 'stripped baryonic dwarf'
galaxies, as they do not possess a dark matter halo.

\item The stripping radius, which is visible in the model galaxy is
in good agreement with an analytical estimate using a Gunn \& Gott
(1972) criterion.

\item In a model galaxy, which moves edge-on through the ICM, the
star formation rate is enhanced as well, however the distribution of
the newly formed stars is different. The new stars are mainly formed
in the central region of the disc.

\item Observations of enhanced SF in merging galaxy clusters and of individual
galaxies with tails of stripped gas support the results presented in
this paper. As a galaxy moves through a real cluster environment, it
encounters gradients in density, temperature, and the gravitational
potential of the cluster and interactions with other galaxies. We
will investigate this complex and degenerate interplay of the
various processes in an upcoming paper.

\end{itemize}

\begin{acknowledgements}
We thank the anonymous referee for fruitful comments which improved
the quality of the paper. The authors also thank Volker Springel for
providing them with GADGET2 and his initial-conditions generator.
Thomas Kronberger is a recipient of a DOC fellowship of the Austrian
Academy of Sciences. The authors further acknowledge the
UniInfrastrukturprogramm des BMWF Forschungsprojekt Konsortium
Hochleistungsrechnen, the ESO Mobilit\"atsstipendien des BMWF
(Austria), the Austrian Science Foundation (FWF) through grants
P18523-N16 and P19300-N16, and the Tiroler Wissenschaftsfonds
(Gef\"ordert aus Mitteln des vom Land Tirol eingerichteten
Wissenschaftsfonds).
\end{acknowledgements}

\end{document}